\documentclass[12pt]{article}

\usepackage{graphicx}
\usepackage[small]{caption}
\usepackage{indentfirst}
\usepackage{afterpage}
\usepackage{longtable}

\begin{document}

\begin{center}
 \textbf{Gravitational stability and dynamical overheating of stellar disks of galaxies.}
\end{center}

 \title{Gravitational stability and dynamical overheating of stellar disks of galaxies.}
\begin{center}
{ Zasov A.V.$^{(1)}$, Khoperskov A.V.$^{(2)}$, Saburova A.S. $^{(1)}$}
\end{center}
\begin{center}
{ \it $^{(1)}$ --- Sternberg Astronomical Institute Moscow State University, Moscow, Russia}
\end{center}
\begin{center}
{\it $^{(2)}$ --- Volgograd State University, Volgograd, Russia}
\end{center}
\bigskip
We use the marginal stability condition for galactic disks and the stellar velocity dispersion data published by different authors to place upper limits on the disk local surface density at two radial scalelengths $R=2h$. Extrapolating these estimates, we constrain the total mass of the disks and compare these estimates to those based on the photometry and color of stellar populations.

The comparison reveals that the stellar disks of most of spiral galaxies
in our sample cannot be substantially overheated and are therefore
unlikely to have experienced a significant merging event in their
 history. The same conclusion applies to some, but not all of
the S0 galaxies we consider.
However, a substantial part of the early type galaxies do show the stellar
velocity dispersion well in excess of the gravitational stability
threshold suggesting a major merger event in the past. We find
dynamically overheated disks among both seemingly isolated galaxies
and those forming  pairs.
The ratio of the marginal stability disk mass estimate to the total galaxy mass
within four radial scalelengths remains within a  range of 0.4---0.8. We
see no evidence for a noticeable running of this ratio with either the
morphological type or color index.

\begin{center}
{\bf 1.Introduction}
\end{center}
The majority of stars in spiral and S0 galaxies are in general concentrated in old stellar
disks with characteristic half-width of several hundreds parsecs. A formation and
dynamical evolution of disks are the matter of hot debates, so the analysis of disk
kinematical characteristics plays a key role (see e.g. Zasov, Sil’chenko (2010) and
references therein). By kinematical characteristics we mean a rotation velocity and
velocity dispersion of old stars, making up the bulk of stellar disk mass, as functions of
radial coordinate in a disk plane. Typical velocity dispersion of old stellar disks equals
to several dozens km/s. In general, it either can reflect the velocity of turbulent motion
of gas which gave the birth to the first stellar generation, or can be the result of the
dynamical heating of disk during its evolution.

Several possible mechanisms of equilibrium disk heating are known, the most effective are
a scattering of stars on massive clouds or stochastic spirals, gravitational interaction
with dark halo subunits, major and minor mergers (see Jenkins, Binney (1990); Binney,
Tremaine (2008); Shapiro et al. (2003)). As far as the velocity dispersion of stars
increases the efficiency of all mechanisms decreases. For initially dynamically cold disk,
another mechanism capable of effectively heating a disk  is the development of
gravitational instability against random perturbations. It leads to the increase of
stellar velocity dispersion both in the plane of a disk, and along the vertical axis
 (through the developing of bending instability). At the early stage of
evolution during the disk mass growth, stellar velocity dispersion might remain fitted to
the level of its marginal stability. If a total disk mass is constant and there exists a
gas disk, forming stars with low velocity dispersion in its plane, a self-regulating
process, supporting stellar disk velocity dispersion in the quasi-stationary station, is
also possible (Betrin, Lin, 1987). In any case, even if gas and young stars are absent,
the minimal velocity dispersion of old stars at given $R$ will be constrained by disk
local marginal stability condition. Numerical simulations of models of initially weakly
unstable 3D disks show the fast passage of a disk into the marginally stable state, after
which the increasing of velocity dispersion practically ceases in the absence of the
heating mechanisms listed above (see e.g. Khoperskov et al. (2003)).

For a thin disk with parameters slowly varying with $R$ the local critical value of radial
velocity dispersion of stars is defined by Toomre criterion:  $c_r=c_T$, where $c_T =
{3.36 G \sigma}/{\kappa}$, $\sigma$ is the disk surface density at given $R$, $\kappa$ is
epicyclic frequency determined by angular velocity $\Omega$ and its radial derivative:
 \begin{equation}
\kappa =2\Omega\cdot \sqrt{(1+(R/2\Omega)\cdot(d\Omega/dR))}.
\end{equation}
A finite disk thickness makes disk more stable, while non-radial perturbations have the
opposite effect (they decrease the stability threshold). The stability condition taking
into account both of these effects can not be expressed in analytic form. If to express
the critical radial velocity dispersion as $(c_r)_{cr}=Q\cdot c_T$, then, as numerical
simulations of 2D and 3D disks show, the parameter $Q$ for a wide range of $R$ lies in the
interval 1.2 --- 2.5  (see e.g. Bottema (1993), Khoperskov et al. (2003)). Note, that the
analytically obtained local stability condition of 2D disk taking into account non-radial
perturbations gives for a flat rotation curve $Q_c\approx 3$ (Polyachenko et al. (1997)).

The assumption of marginally stable disks makes it possible to put constraints on its
surface density:
\begin{equation}\label{Eq-cr-QToomre}
\sigma (R) = \frac{\kappa (R) c_r (R)}{3.36 G Q(R)},
\end{equation}
and hence on a total disk mass therefore.  It enables to obtain relative masses of disk
and spheroidal components (bulge and dark halo) (Bottema (1993), Bottema (1997), Tyurina
et al. (2001), Zasov et al. (2006)) and to estimate local thickness of stellar disk (see
f.e. Khoperskov et al., 2010), which allows in turn to find a volume density of gas layer
in disk plane if it exists in a galaxy (Zasov, Abramova (2006), Kasparova, Zasov (2008),
Zasov, Abramova (2008)).

However the question of how far are the real disks from marginally stable state still
remains open. There are two indirect evidences of closeness of stellar velocity dispersion
to $(c_r)_{cr}$ for significant fraction of disk galaxies. First, disk mass-to-light
ratios, which follow from the stability condition are in good agreement with the estimates
based on stellar population evolution photometrical models (see e.g. Bottema (1997), Zasov
et al. (2004)). Second, the assumption of disk marginal stability enables to explain a
positive correlation of relative disks thicknesses and their relative masses within the
optical borders (Zasov et al. (2002)). Note, however, that both numerical simulations
(Sotnikova, Rodionov (2006), Khoperskov et al. (2010)) and the analysis of observations of
edge-on galaxies (Mosenkov et al. (2010), Bizyaev, Mitronova (2009), Bizyaev (2010)) show
that this correlation is rather sparse especially for galaxies with small bulges.
Moreover, numerical dynamical models by far are not always compatible with the assumption
of disk marginal stability (see, e.g., Zasov et al. (2008)).

A difficulty of estimation of the ratio of observed velocity dispersion $c_{obs}$ to its
critical value $(c_r)_{cr}$ lies first of all in the difficulty of measurement of
$c_{obs}$ at sufficiently large galactocentric distances and in addition is due to a
number of factors hard to be taken into account even in the axissymmetrical galaxy model.
Among the latter factors are the variation of critical Toomre parameter with radius, the
ratio of radial to vertical velocity dispersion and the account of asymmetric drift in the
estimation of disk angular velocity especially for the early type galaxies.

The main goal of this work is to find out how far the observed values of velocity
dispersion at given $R$ are from the expected ones for marginally stable disks for
different type galaxies. For this purpose, we analyze the velocity dispersion data,
available in the literature, for a chosen galactocentric distance $R=2h$, where $h$ is the
radial disk scalelenght. At this radius a disk contribution to the rotation curve is
nearly maximal. At higher $R$, measurements of velocity dispersion are less reliable and
also the influence of external factors dynamically heating a disk may be more important.
In turn, at lower $R$ it is often hard to separate the contributions of bulge and disk
stars to the measured values of $c_{obs}$, and to take into account the presence of a bar
if it occurs in a galaxy.  It is worth noting that, as numerical 3D N-body simulations
show for different parameters of components (disk, bulge and halo), the spread of values
of $Q(R)$ is minimal at $R\approx2h$. Following Khoperskov et al. (2003) here we assume
$Q(2h)\approx 1.5$.
\medskip
\begin{center}
{\bf 2.  The sample and the method we use}
\end{center}
 \medskip
To estimate the upper limits of disk masses corresponding to the marginal stability
condition we have chosen objects for which sufficiently extensive radial distributions of
stellar velocity dispersion and rotation curves were available in the literature (121
objects). Using these data we estimated the threshold value of surface density
corresponding to the gravitational stability at $R=2h$ from the equation (2). Radial
velocity dispersion of stars $c_r$, if it was not given in the cited literature, was
determined from $c_{obs}(R)$ measured along the major axis of a galaxy:
\begin{equation}\label{Eq-c-obs}
c_{obs}^2 = c_{\phi}^2 \sin^2 i + {c_z}^2 \cos^2 i \,,
\end{equation}
which leads to:
\begin{equation}\label{Eq-cr-cobs}
c_r=c_{obs}\cdot \left[(c_{\phi} / c_r)^2 \sin^2 i + (c_z/c_r)^2 \cos^2 i\right]^{-1/2}
\,.
\end{equation}
where $c_{\phi}$ and $c_z$  are azimuthal and vertical components of velocity dispersions
correspondingly, and  $i$ is the inclination of a disk. Radial and azimuthal velocity
dispersions are interconnected by Lindblad formula $c_{\phi}/c_r=\kappa/ 2\Omega$
following from the epicyclic approximation. Direct measurements of  $c_z/c_r$ ratio in
galaxies show that in majority of cases it lies in the interval 0.4---0.7 (Shapiro et al.
(2003)). In current work it was taken to be 1/2.

In most cases rotation curve $V(R)$ has a plateau at the distance of two disk radial
scalelenghts, so the $\kappa/\Omega $ ratio can be taken as $\sqrt 2$. For the increasing
or decreasing rotation curves we applied the corresponding corrections. If only the
stellar rotation curve $V_*(R)$ was available for a galaxy, the approximate asymmetric
drift correction was introduced following Neistein et al. (1999). Note, that for
$c_r/V_c>0.5$ this correction is inevitably very crude.

A list of galaxies with available measurements of stellar disk velocity dispersion and
their observed and calculated properties is given in Table 1. For galaxies with
line-of-sight velocity exceeding 800~km/s the adopted distances correspond to the Hubble
constant $H_0=75$~km/s/Mpc, (except Virgo cluster galaxies for which we took $D=17 $~Mpc).
For several nearby galaxies with lower line-of-sight velocity (NGC 3198, NGC 6503, IC750)
we used the distance modulus given in Hyperleda database. The disk inclination $i$, color
indices corrected for galactic extinction, inclination and redshift, and line-of-sight
velocities were also taken from this database. For the cases when the inclinations taken
from Hyperleda differed significantly from those given in the original sources referred in
the Table, the preference was given to the later ones.

After the estimation of surface density at $R=2h$, it is easy to find a total disk mass:
\begin{equation}\label{Eq-Mdisk}
M_d = 2\pi h^2\, \sigma(2h)\, e^2 \,.
\end{equation}
In this work a disk photometric radial scalelenght and the scalelenght of surface density
are assumed to be equal. In some cases this can be quite rough approximation, but the
introduced  uncertainty is not so high in comparison with other error sources: a change of
$h$ by $\pm$ 30\% for a fixed adopted value $\sigma(2h)$ leads to the change of $M_d$ by
+7\%, -15\%.

To compare the obtained disk masses with the photometric estimates we calculated disk
mass-to-light ratios in $B$ band $(M/L_B)_d$. For Sb and later type galaxies where the
bulge contribution to the luminosity is low, we neglected the difference between the total
and disk luminosity. For Sa---S0 galaxies we have taken into account the contribution of
bulge. To do this we took a disk luminosity estimates or disk de-projected surface
brightness at $R=2h$ from de-projected radial surface brightness profiles available in
the literature. In the absence of such data a disk luminosity was calculated from the total
luminosity using bulge-to-disk luminosity ratio $B/D$, average for given morphological
type (taken from Graham (2001)): $L_{d}={L_{tot}}/{((B/D)+1)}$. Disk color index was
obtained in this case from the total color index:
 \begin{equation}
(B-V)_{d}=2.5\cdot \lg\left(10^{(0.4\cdot(B-V)_{tot})}-B/T\cdot 10^{(0.4\cdot
(B-V)_{bulge})}\right) \,.
\end{equation}
Bulge color index $(B-V)_{bulge}$ of early type galaxy was taken to be 0.8, if the total
color index of galaxy did not exceed 0.8. Otherwise (for very red galaxies), a color index
of a disk was accepted to be equal to the total one.
\begin{center}
{ \bf 3. Results}
\end{center}
Surface density and disk mass estimates together with other data and the references to the data sources are listed in Table 1. It contains:\\
({\it 1}) Name.\\
({\it 2}) Morphological type.\\
({\it 3}) Radial disk scalelength $h$ used in the paper.\\
({\it 4}) Ratio of radial velocity dispersion to circular velocity at $R=2h$. \\
({\it 5}) Disk surface density corresponding to the marginal stability condition at $R=2h$.\\
({\it 6}) Mass-to-light ratio of a disk, marginally stable at $R=2h$ ($B$-band).\\
({\it 7}) Ratio of mass of the disk, marginally stable at $R=2h$, to the total mass $M_t$ within $R=4h$,
where $M_t = 4 V_c^2 h / G$ and $V_c$ is circular velocity. Cases of $M_{d}/M_t >1$  indicate a  significant dynamical overheating of a disk. \\
({\it 8}) Velocity dispersion and rotation velocity data sources.\\

Diagrams ``$(M/L_B)_{d}$---$(B-V)_0$'' for stellar disks are shown in Fig. 1,
where disks masses $M_d$ are obtained from the gravitational stability condition at
$R=2h$.\footnote {Color indices are available not for all considered galaxies, so the
number of points in the diagram is not equal to the number of galaxies listed in Table 1.}
For the Galaxy the mean color index for $Sbc$ type was used (Buta et al., 1994). A
straight line reproduces model relation obtained by Bell, de Jong (2001) for old stellar
systems with different present-day star formation rate (SFR) using modified (bottom-light)
Salpeter initial mass function (IMF). It is worth noting that the light extinction inside
the galaxies that may play important role for disks with high content of dust decreases
the luminosity of a galaxy and simultaneously increases its color index shifting the
points practically along the theoretical correlation.
\begin{figure} [h!]
\includegraphics[width=12cm,keepaspectratio]{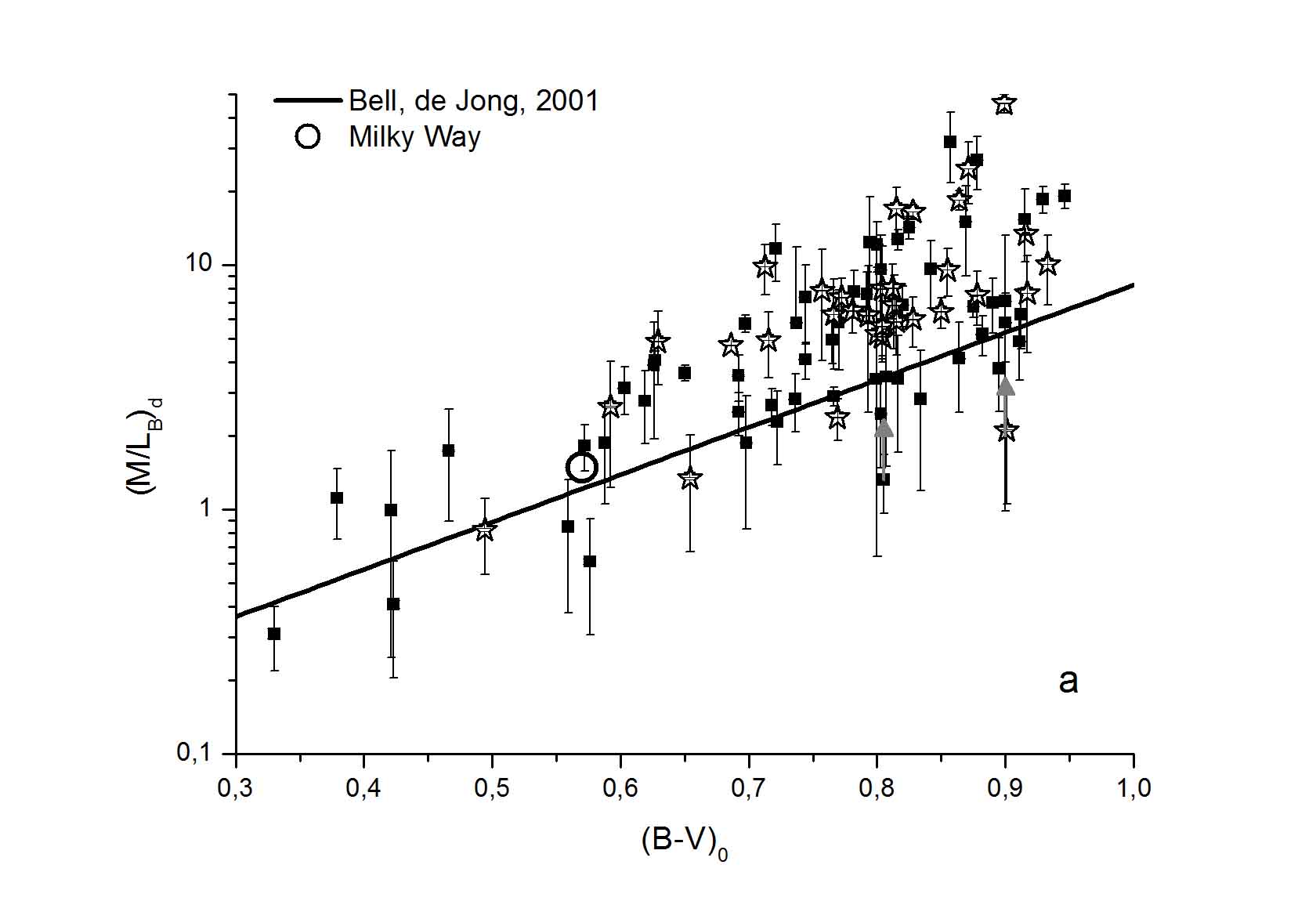}
\includegraphics[width=12cm,keepaspectratio]{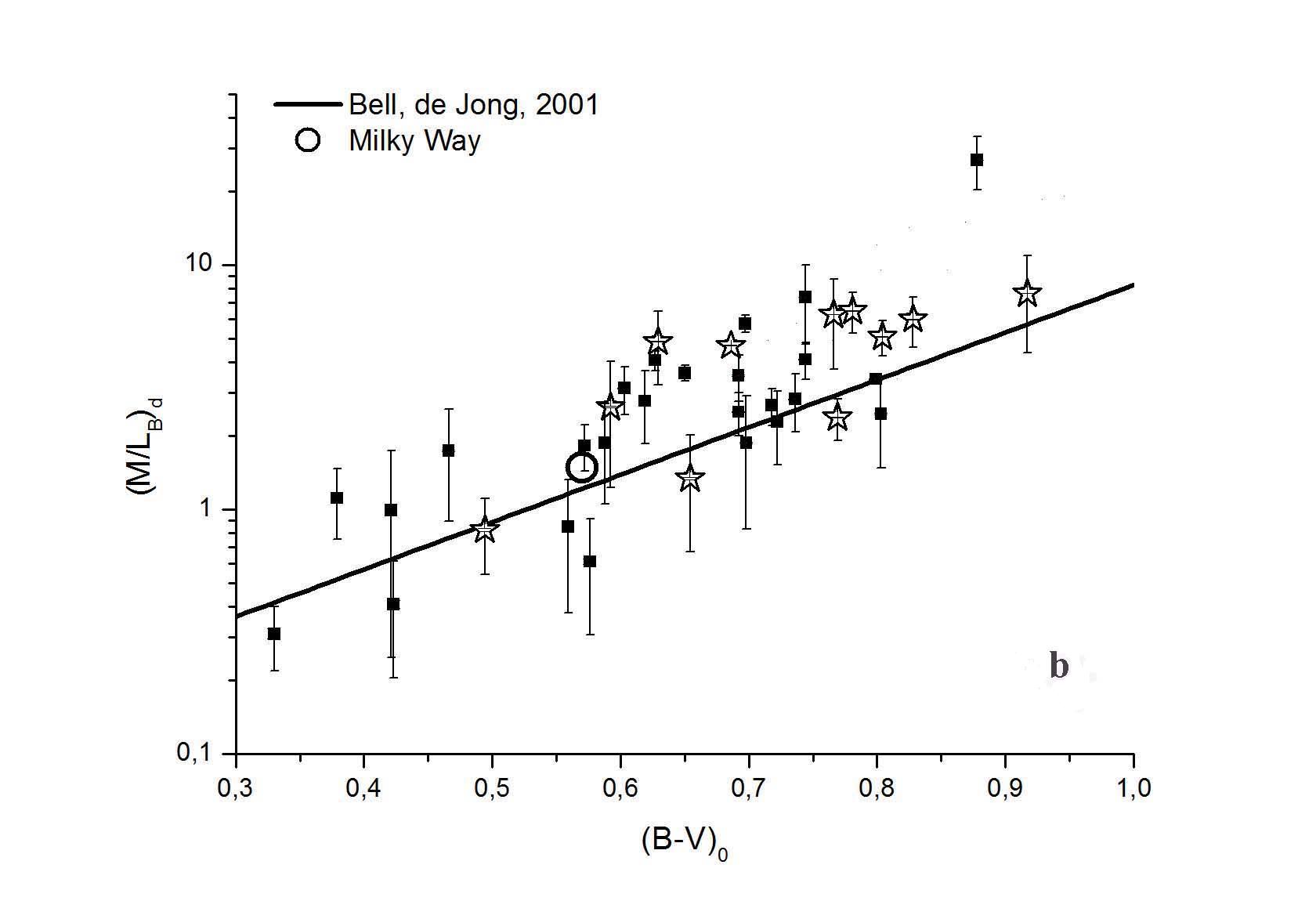}
\caption{Diagram ``$(M/L_B)_d$---$(B-V)_0$'', plotted for (a) the entire sample of
galaxies and (b) for the galaxies of types later than $S0/a$. Milky Way is marked by open
circle, pair members are shown by asterisks. Straight line corresponds to model relation
obtained for old stellar systems with different history of star formation by Bell, de Jong
(2001). Here and in other figures the error bars correspond to the errors of velocity
dispersion  data used to estimate the disk masses.}
\end{figure}

In Fig. 1a the entire sample of galaxies with known color indices is shown; pair members
are marked by asterisks. Here and in the other figures the error bars correspond to the
errors of velocity dispersion used to calculate the disk local surface density. The
uncertainty of parameter $Q_c$ is not included in the error bars (according to the
numerical simulations it reaches at least 20\%). The same diagram as in Fig. 1a, but after
the exclusion of S0-a---S0 galaxies, is shown in Fig. 1b.  Though the scatter of points in
these diagrams is large ($\sim 0.3~dex$), it is compatible with the errors of individual
mass estimates. Therefore one can conclude that there is a general agreement between
$(M/L_B)_{d}$ estimates based on marginal stability condition and those based on the
stellar population evolution modeling. It is remarkable that most of the galaxies, which
significantly deviate from model relation have a red color $(B-V)_0>0.7$. At least half of
these galaxies are above the straight line, that is most of them have disks with
significant dynamical overheating. Note, that there are both paired and isolated systems
among these galaxies. Pair members occur also among the late type galaxies in Fig. 1b (11
objects), but the presented statistical data are not enough to claim their systematical
difference from the isolated systems. Anyway, pair, group and cluster membership is not
always connected with the stellar velocity dispersion exceeding the level needed for
marginal gravitational stability. It seems that only a strong gravitation perturbation can
be a cause of disk dynamical overheating. This conclusion also follows, for example, from
the higher mean thickness of stellar disks in interacting edge-on galaxies (Reshetnikov,
Combes (1997)).

The position of our Galaxy in the diagrams is shown by the open circle. For our Galaxy
$(M/L_B)_d$ ratio was calculated using the local values of surface density and brightness
measured for solar vicinity and accepting $h=3$ kpc. The surface brightness data in $B$-band was taken from Portinari et al. (2005)..
Note, that the value of surface density in the Solar neighborhood obtained on the basis of
marginal stability condition is in a good agreement with the direct estimates (Korchagin
et al. (2003), Holmberg, Flynn (2004), Kuijken, Gilmore (1991)). This gives evidences of
marginal stability of disk of Galaxy at least at the galactocentric distance of a few
radial scalelenghts. This conclusion conforms with the idea of the absence of major merger
events in the history of our Galaxy (see e.g. Wyse, 2009)).

In Fig. 2a the ratio of radial velocity dispersion to circular velocity $c_r/V_c$ is
correlated with morphological type of galaxies. Type $t$ is coded as de Vaucouleurs
numerical type: spiral galaxies correspond to the interval $t$=1 ($Sa$) --- 7 ($Sd$),
higher values of $t$ represent irregular galaxies, and  $t=-1...-3$ are related to S0
galaxies. The correlation between $c_r/V_c$ and type is rather weak, moreover, if to
exclude S0 galaxies, it almost disappears. The same conclusion  may be done for the
correlation between $c_r$  and $V_c$ at the fixed radius $R=2h$ (Fig. 2b). It means that
it is impossible to get a reliable estimation of disk velocity dispersion by the indirect
way, from the type of a galaxy or its rotation velocity, as it was for example suggested
by Bottema (1993).

The conclusion that most of the galaxies with dynamically overheated disks are among the
red galaxies (Fig. 1a) is confirmed by Fig. 3, where color index is compared with the
relative disk mass $M_d/M_t$ within the optical radius $R=4h$. As above, here $M_d$
corresponds to disks, which are assumed to be marginally stable at least at $R=2h$, and
total mass of a galaxy was calculated as $M_t = 4h\cdot V_c^2/G$. \footnote{Note that this
simple formula for a total mass, is, strictly speaking, correct for spherically symmetric
distribution of density and may underestimate $M_{d}/M_t$ by 20-30\% $M_{d}/M_t$ for a
disk galaxy which does not possesses a massive bulge or halo.}
\begin{figure} [h!]
\includegraphics[width=12cm,keepaspectratio]{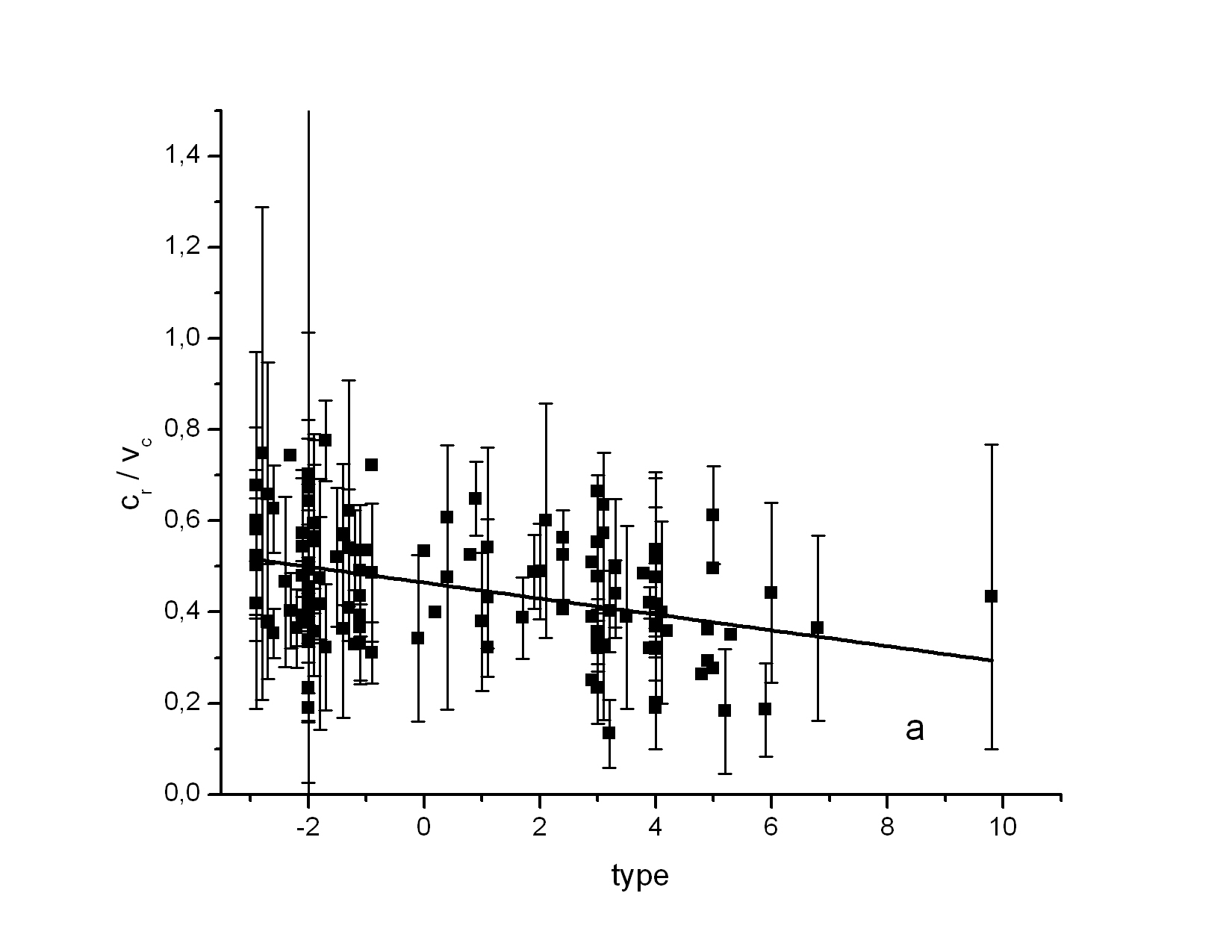}
\includegraphics[width=12cm,keepaspectratio]{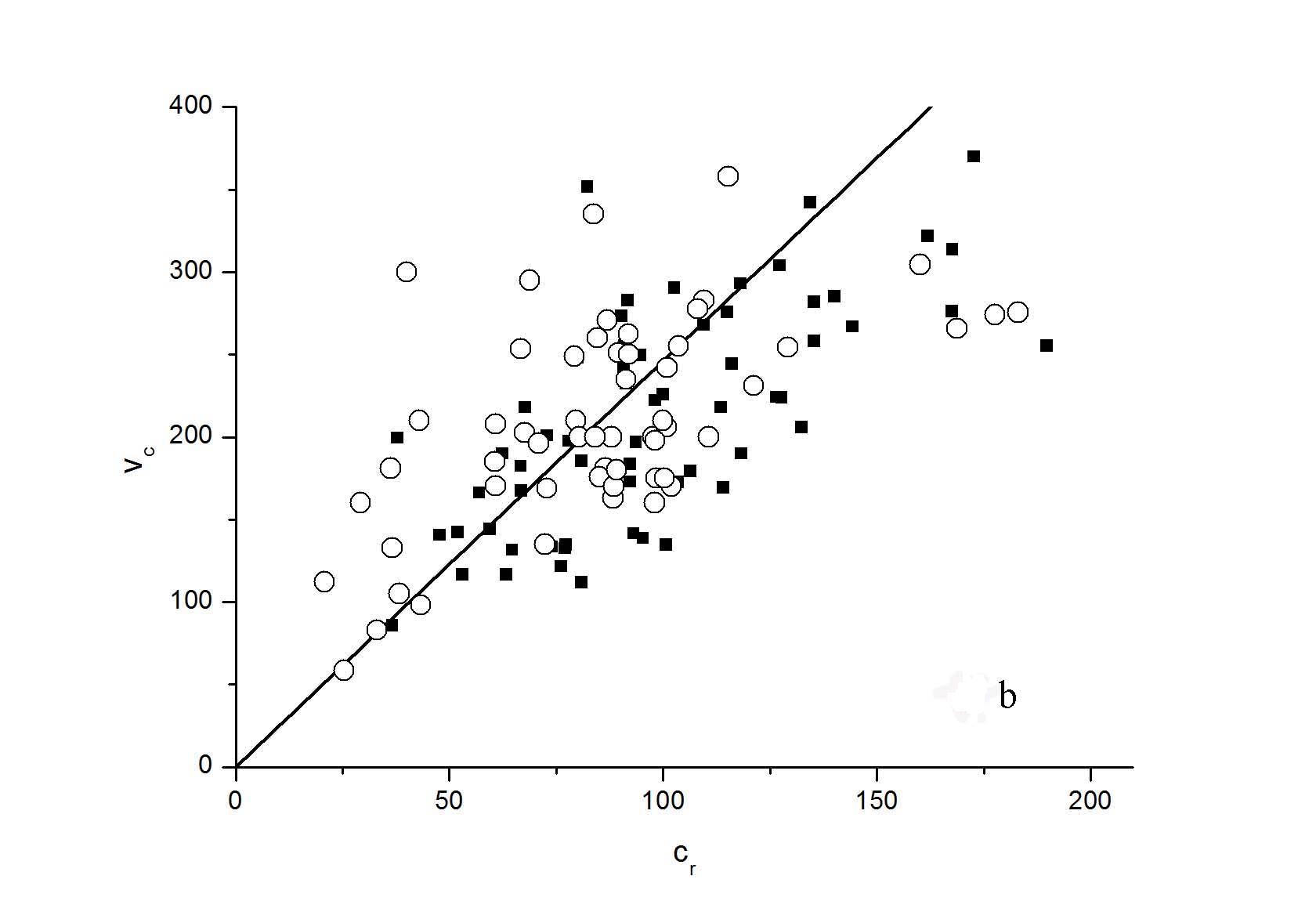}

\caption{(a) Diagram ``$c_r/V_c$ --- morphological type''. (b) Diagram ``$V_c$---$c_r$''.
Galaxies with types later than Sa are marked by open circles; a straight line corresponds
to the mean value of $c_r/V_c$.}
\end{figure}

\begin{figure} [h!]
\includegraphics[width=12cm,keepaspectratio]{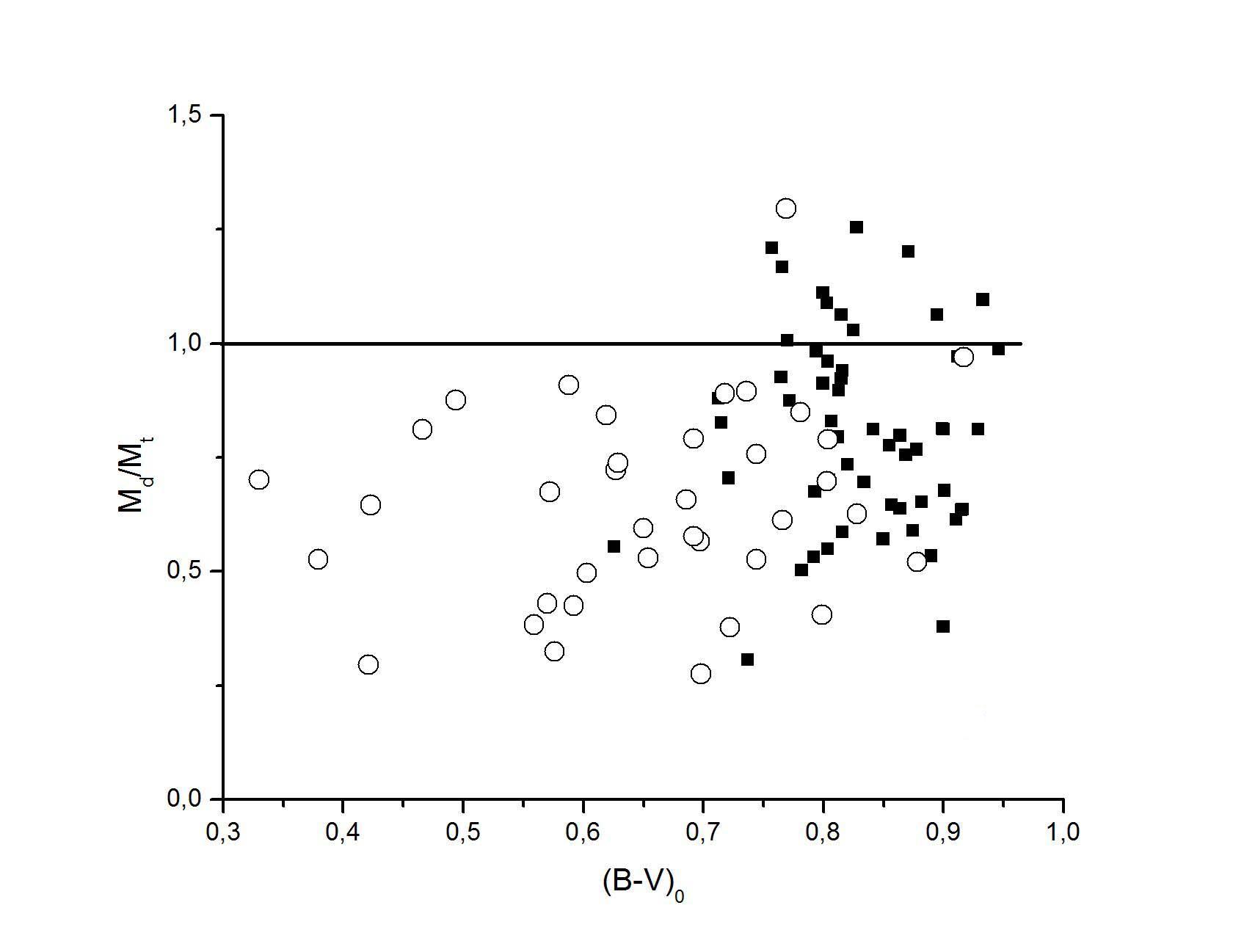}

\caption{Diagram ``$M_{d}/M_t$ --- $(B-V)_0$''. A  horizontal line marks $M_{d}/M_t=1$.
Points above this line correspond to the galaxies with certainly overheated disks. Open
circles represent the galaxies with types later than Sa. Typical errors of $M_{d}/M_t$
caused by the uncertainty of stellar velocity dispersion estimates are about 40\%.}
\end{figure}

As far as $M_d$ we found is in general case the upper limit of mass of a disk, the values
$ M_d/M_t \ge 1 $ obtained for some galaxies (see Fig. 3) have no physical sense and
indicate unambiguously the significant overestimation of disk masses, or, in other words,
the disk overheating. Almost all of these objects are lenticular galaxies with a high
color indexes $(B-V)_0>0.7$.

  Lenticular galaxies differ from other galaxies of the sample by their distributions of both $M_{d}/M_t$ ratios (Fig. 3)
  and ``marginal'' surface densities at  $R=2h$ (Fig. 4), being shifted to higher values of $M_{d}/M_t$ and densities.
  These differences are statistically significant: according to the Wilcoxon criterion, the level of significance $\alpha=0.99$.
From the distribution of $M_{d}/M_t$ (Fig. 3) it follows that for most of galaxies disk to
total mass (inside $R=4h$) ratios lay in the interval 0.4---0.8, independently of color and
morphological type (if to exclude galaxies with apparently overheated disks, for which
these ratios remain uncertain).
\begin{figure} [h!]
\includegraphics[width=14cm,keepaspectratio]{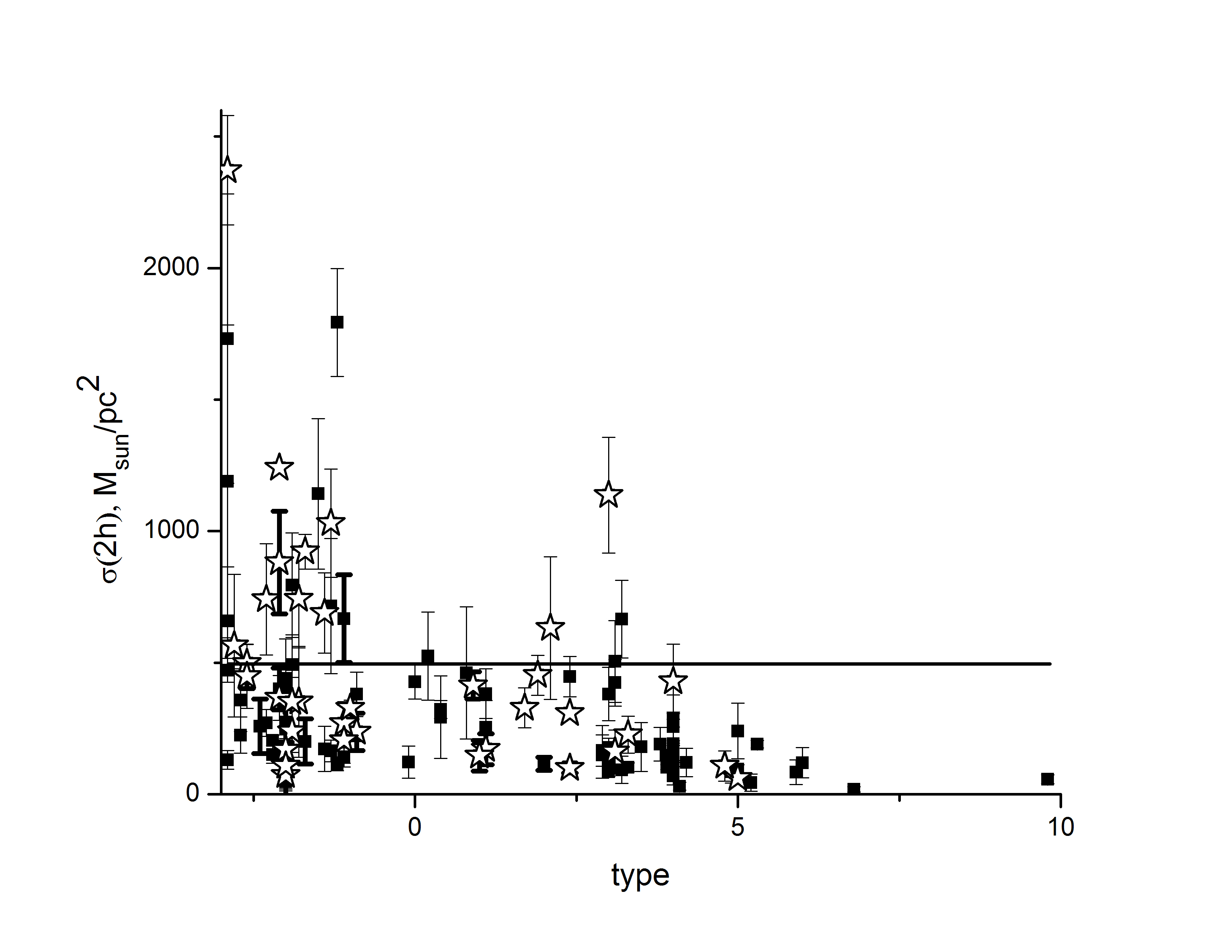}
 \caption{Diagram ``$\sigma(2h)$ --- morphological type''. Pair members are marked by asterisks.
 Straight horizontal line separates galaxies with $\sigma(2h)=500 M_{sun}/$pc$^2$ level, which definitely evidences the  overestimation of their mass.}
\end{figure}

If to consider the disky galaxies as systems with marginally stable disks, one can plot
the relation $M_d$---$V_c$ (baryonic Tully-Fisher relation), connecting the most important
parameter of stellar disk (a mass) with the parameter determined mostly by massive halo
(rotation velocity). This diagram is shown in Fig. 5 (galaxies with $M_{d}/M_t>1$ are
excluded). The relation obtained by McGaugh (2005) for a large sample of galaxies, disk
masses of which were estimated from their luminosities and colors, is also shown by
straight line for comparison. From this figure it follows that disk masses estimated from
the marginal stability condition show the same relation as masses obtained from photometry
and evolution models, which confirms the absence of systematical difference between these
evaluations.

\begin{center}
{\bf 4. Conclusions}
\end{center}
As it was shown above, if to assume the marginal stability of inner parts of disks of
spiral galaxies, one can estimate local values of disk surface densities and as the next
step --- the absolute and relative values of their total masses or, strictly speaking, the
upper limits of masses for stable equilibrium disks. These estimates in most cases appear
to be in a good agreement with the values obtained from photometric data on the basis of
stellar population evolution models (Fig. 1b, Fig. 5). It allows to admit that the disks
of most spiral galaxies are close to the marginally stable state.

Evidently, mass and density estimates presented in this paper for any given galaxy remain
rather rough, mainly  due to large errors of radial velocity dispersion $c_r$, and, to a
lesser degree, of radial scalelength $h$ values. For some of the galaxies the resulting errors
exceed a factor of 2 (see Fig. 1). Thus to obtain more reliable estimates for individual
galaxies one have to construct dynamical models consisted of several components which
reproduce the radial profiles of observed rotation velocity and velocity dispersion
$c_{obs}$ of disk stars. Still, the estimation of disk masses for galaxies of different
types presented in this paper allows to conclude that there is no significant systemic
dynamical overheating of disks in most of galaxies, except lenticulars, where the velocity
dispersion often exceeds the threshold needed for marginal stability. It follows that many
spiral galaxies  have not suffered a strong gravitation perturbations or mergings after
the formation of bulk of their stars. This result creates a certain problem in the
generally accepted theory of galaxies formation in a gravitational well of dark halo
(see e.g. the discussion in Zasov, Silchenko, 2010). The conclusion of marginal stability
of disks, however, may not concern their outermost regions, which require an individual
analysis and often may be dynamically overheated.

It is essential, that the excessively high velocity dispersion at $R=2h$ usually takes
place in galaxies which, if to judge by color indexes, have a weak star formation (color
index $(B-V)_0 > 0.7$). Most of them belong to $S0/a-S0$ types.  For these galaxies almost
total absence of young stars and gas parallel with high stellar velocity dispersion
apparently resulted from at least one merger with massive satellite in their history. This
event could both increase the velocity dispersion of stars and decrease the content  of
cold gas evidently as the result of intensive star formation burst triggered by merging.
Gas partially may transform into stars, partially could be ejected from the disk due to
the activity of a large number of massive stars.

Nevertheless, as it follows from Figs 1a and 3, some  ``red'' galaxies  still appear to
keep their disks in the state close to marginal stability despite a high color index. It
supports the idea of the existence of different mechanisms responsible for the
transformation of spiral galaxies into lenticulars.\\

This work was supported by grants RFBR 07-02-00792, 09-02-97021.
 \pagebreak
\begin{figure} [h!]
\includegraphics[width=14cm,keepaspectratio]{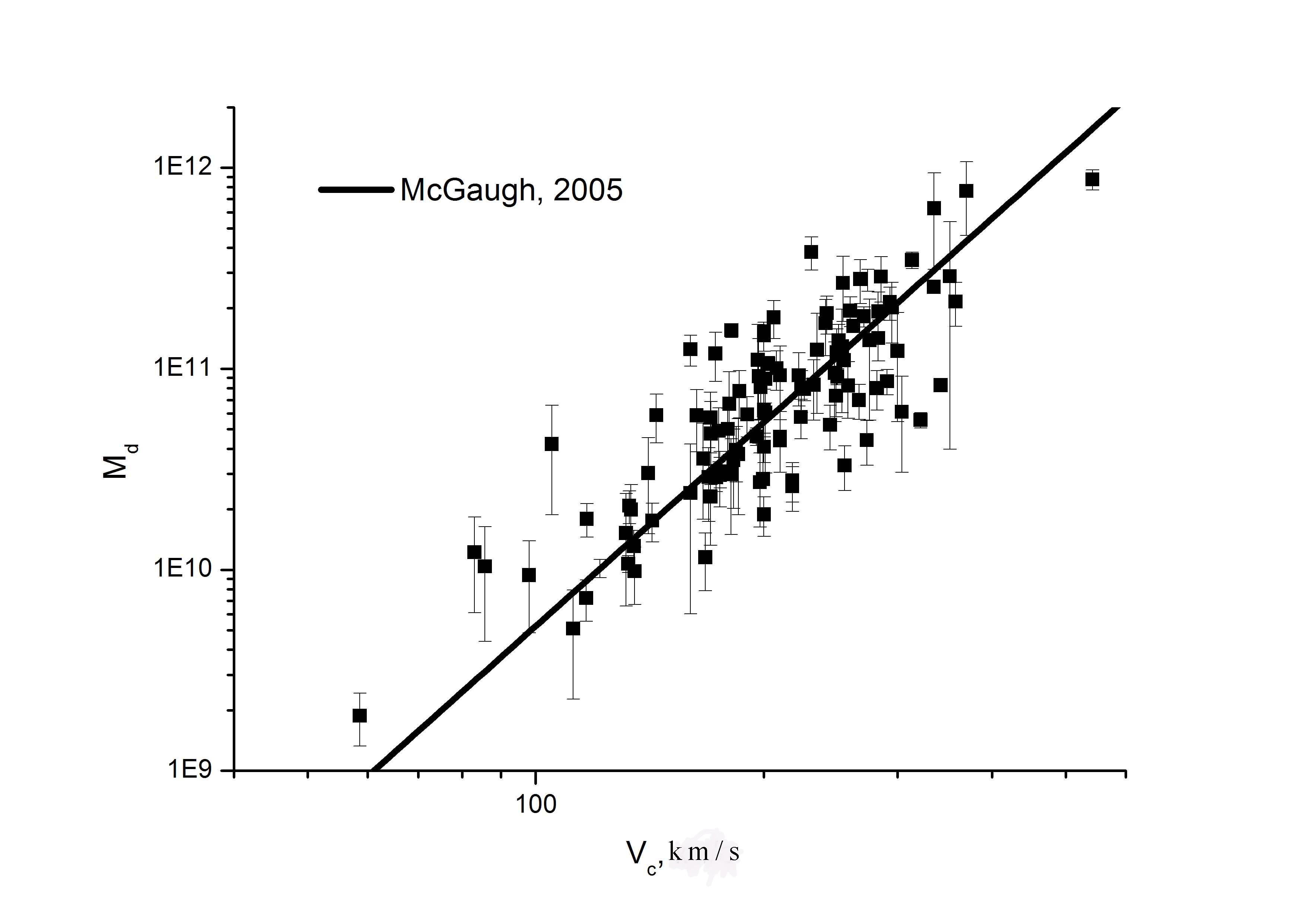}
\caption{Baryonic Tully-Fisher relation between the disk mass and rotation velocity.
Straight line corresponds to the relation presented by McGaugh (2005) for spiral galaxies
where disks masses were found from their luminosity and color .}
\end{figure}


\pagebreak
\begin{longtable}{|c|c|c|c|c|c|c|c|c|c|}
\caption{Main results}\label{Tabl1} \\
\hline {\it 1} & {\it 2}& {\it 3} &{\it 4}&{\it 5}& {\it 6}&{\it 7}&{\it 8}\\
\hline Name & Type& $h$    & $c_r/V_c$ & $\sigma(2h)$        &$(M/L_B)_d$&$M_d/M_t$& Ref \\
                 &    &arcsec.&           & $M_{sun}/$pc$^2$&           &         &\\\hline
\endfirsthead
\caption{Continuation }\label{Tabl1} \\
\hline {\it 1} & {\it 2}& {\it 3} &{\it 4}&{\it 5}& {\it 6}&{\it 7}&{\it 8}\\
\hline Name & Type& $h$    & $c_r/V_c$ & $\sigma(2h)$        &$(M/L_B)_d$&$M_d/M_t$& Ref \\
                 &    &arcsec.&           & $M_{sun}/$pc$^2$&           &         &\\\hline
\endhead
\hline
\endfoot

MW  &   SBbc    &   3 [kpc]   &   0.2 &   97.3    &   1.91    &   0.33    &   \cite{73}   \\
NGC1023 &   E-S0    &   61.5    &   0.38    &   225 &   4.87    &   0.61    &   \cite{8}    \\
NGC1068 &   Sb  &   55.6    &   0.36    &   177 &   2.5 &   0.58    &   \cite{1}   \\
NGC1566 &   SABb    &   35.7    &   0.2 &   69.7    &   0.61    &   0.32    &   \cite{3}, \cite{31}   \\
NGC2460 &   Sab &   15  &   0.49    &   452 &   5.09    &   0.8 &   \cite{1}   \\
NGC2549 &   S0  &   20.2    &   0.5 &   354 &   9.62    &   0.81    &   \cite{8}   \\
NGC2613 &   Sb  &   57.3    &   0.13    &   92.9    &   1.87    &   0.28    &   \cite{3}, \cite{32}   \\
NGC2768 &   S0   &   44.8    &   0.6 &   130 &   6.27    &   0.97    &   \cite{8}   \\
NGC2775 &   Sab &   35  &   0.39    &   328 &   6   &   0.63    &   \cite{1}   \\
NGC2962 &   S0-a    &   22.3    &   0.39    &   264 &   15.4    &   0.64    &   \cite{6}    \\
NGC2985 &   Sab &   30  &   0.41    &   307 &   4.69    &   0.66    &   \cite{13}  \\
NGC3198 &   Sc  &   58  &   0.18    &   44.1    &   0.99    &   0.3 &   \cite{3}, \cite{33}    \\
NGC338  &   Sab &   18.3    &   0.49    &   116 &   3.53    &   0.79    &   \cite{11}   \\
NGC3489 &   S0-a    &   19.4    &   0.62    &   715 &   5.82    &   1.01    &   \cite{8}    \\
NGC3630 &   S0  &   8.3 &   0.52    &   1140    &   8.23    &   0.84    &   \cite{6}, \cite{7}     \\
NGC3949 &   Sbc &   20  &   0.54    &   286 &   1.11    &   0.53    &   \cite{19}  \\
NGC3982 &   SABb    &   10.2    &   0.4 &   666 &   2.14    &   0.65    &   \cite{2}    \\
NGC4030 &   Sbc &   26.2    &   0.32    &   257 &   2.73    &   0.51    &   \cite{1}    \\
NGC4143 &   S0  &   15.4    &   0.36    &   794 &   8.27    &   0.58    &   \cite{7}    \\
NGC4203 &   E-S0    &   17.2    &   0.66    &   358 &   3.76    &   1.06    &   \cite{7}    \\
NGC4251 &   S0  &   22.6    &   0.56    &   492 &   12.1    &   0.91    &   \cite{8}    \\
NGC4419 &   SBa &   19  &   0.43    &   254 &   2.46    &   0.7 &   \cite{5}   \\
NGC4578 &   S0  &   36.4    &   0.34    &   72.4    &   7.98    &   0.55    &   \cite{7}    \\
NGC4649 &   S0   &   29.6    &   0.68    &   1730    &   10.1    &   1.1 &   \cite{8}    \\
NGC470  &   Sb  &   13.6    &   0.33    &   162 &   1.34    &   0.53    &   \cite{5}    \\
NGC4753 &   S0-a    &   34.8    &   0.36    &   172 &   3.42    &   0.59    &   \cite{8}    \\
NGC488  &   Sb  &   40  &   0.25    &   149 &   3.41    &   0.4 &   \cite{12}  \\
NGC5247 &   SABb    &   40  &   0.4 &   30  &   0.41    &   0.65    &   \cite{3}   \\
NGC5273 &   S0  &   18.3    &   0.56    &   238 &   6.82    &   0.9 &   \cite{7}   \\
NGC5440 &   Sa  &   15.7    &   0.65    &   414 &   16.9    &   1.05    &   \cite{11}   \\
NGC5533 &   Sab &   34.4    &   0.53    &   100 &   6.48    &   0.85    &   \cite{11}   \\
NGC584  &   S0   &   8.5 &   0.42    &   1190    &   2.1 &   0.68    &   \cite{8}   \\
NGC5866 &   S0-a    &   21  &   0.54    &   1031    &   7.35    &   0.87    &   \cite{8}    \\
NGC598* &   Sc  &   319.2   &   0.44    &   120 &   1.74    &   0.81    &   \cite{15}, \cite{16}    \\
NGC6340 &   S0-a    &   28  &   0.48    &   293 &   12.4    &   0.98    &   \cite{14}   \\
NGC6503 &   Sc  &   40  &   0.19    &   84  &   0.85    &   0.38    &   \cite{3}   \\
NGC7469 &   Sa  &   8.3 &   0.54    &   171 &   0.83    &   0.88    &   \cite{11}  \\
IC750   &   Sab &   14.2    &   0.6 &   631 &   7.65    &   0.97    &   \cite{4}   \\
NGC7743 &   S0-a    &   18.7    &   0.49    &   237 &   4.92    &   0.83    &   \cite{11}   \\
NGC7782 &   Sb  &   18.3    &   0.23    &   99.2    &   2.29    &   0.38    &   \cite{5}    \\
ES0 288-G25 &    Sbc                &   18.4    &   0.36    &   121 &   3.28    &    0.61   &   \cite{62}   \\
ES0 435-G14 &    Sc                 &   17.6    &   0.28    &   60.9    &   1.6 &    0.47   &   \cite{62}   \\
ES0 435-G25 &    Sc                 &   47.35   &   0.29    &   84.9    &   3.13     &  0.5 &   \cite{62}   \\
ES0151-004  &    S0                 &   10.6    &   0.38    &   138 &   9.51    &    0.61   &   \cite{59}   \\
ES0189-7    &   SABb    &   31.1    &   0.42    &   102 &   4.08    &   0.72    &    \cite{20} \\
ES0240-011  &    Sc                 &   29.5    &   0.26    &   107 &   2.63    &    0.43   &   \cite{59}   \\
ES0311-012  &    S0-a               &   21.4    &   0.53    &   427 &   3.23    &    0.86   &   \cite{59}   \\
ES0358-g006 &   S0   &   9.2 &   0.58    &   472 &   12.7    &   0.94    &   \cite{47}   \\
ES0446-17   &   SBb &   19.4    &   0.44    &   103 &   3.86    &   0.71    &   \cite{20}   \\
ES0450-20   &   SBbc    &   17.3    &   0.37    &   281 &   4.36    &   0.6 &   \cite{20}   \\
ES0514-10   &   SABc    &   27.1    &   0.61    &   95.5    &   3.49    &   0.99     &  \cite{20}  \\
ES0597-036  &    S0-a               &   11  &   0.41    &   165 &   14  &   0.69     &  \cite{59}  \\
IC1963  &   S0  &   15.8    &   0.36    &   151 &   5.12    &   0.59    &   \cite{47}   \\
IC4767  &    S0-a               &   16  &   0.44    &   141 &   11.6    &   0.7 &    \cite{59} \\
IC4937  &    Sb                 &   21  &   0.48    &   84.5    &   8.86    &    0.81   &   \cite{59}  \\
IC5096  &    Sbc                &   26.6    &   0.32    &   148 &   5.31    &    0.52   &   \cite{59}  \\
IC5249  &   SBcd    &   40.1    &   0.37    &   18.6    &   1.6 &   0.59    &    \cite{80}    \\
m31*    &   Sb  &   1548    &   0.33    &   121 &   4.1 &   0.53    &   \cite{28}  \\
NGC0524 &   S0-a    &   19.2    &   0.61    &   1790    &   19.1    &   0.99    &    \cite{6}  \\
NGC0980 &   S0  &   12.75   &   0.32    &   170 &   8.07    &   0.52    &   \cite{5}    \\
NGC1032 &    S0-a               &   26.6    &   0.61    &   322 &   14.3    &    1.03   &   \cite{59}  \\
NGC1052 &   E-S0   &   17.3    &   0.52    &   658 &   5.81    &   0.81    &   \cite{8}    \\
NGC1167 &   S0  &   24.63   &   0.47    &   258 &   15  &   0.76    &   \cite{38}, \cite{37}   \\
NGC1175 &   S0-a    &   12.9    &   0.49    &   269 &   8.04    &   0.79    &   \cite{46}   \\
NGC128  &   S0  &   24.5    &   0.23    &   141 &   7.12    &   0.38    &   \cite{45}   \\
NGC1316 &   S0  &   49.4    &   0.78    &   922 &   16.5    &   1.25    &   \cite{47}   \\
NGC1375 &   S0  &   14.8    &   0.54    &   366 &   9.84    &   0.88    &   \cite{47}   \\
NGC1380 &   S0  &   36.3    &   0.4 &   271 &   5.21    &   0.65    &   \cite{47}  \\
NGC1380A    &   S0  &   19.8    &   0.45    &   104 &   6.87    &   0.74    &   \cite{47}   \\
NGC1381 &   S0  &   20.8    &   0.36    &   204 &   6.76    &   0.59    &   \cite{47}   \\
NGC1461 &   S0  &   22  &   0.44    &   403 &   18.6    &   0.81    &   \cite{49}  \\
NGC1886 &    Sbc                &   24.3    &   0.48    &   190 &   6.1 &    0.78   &   \cite{59}   \\
NGC2273 &   SBa &   29.7    &   0.38    &   146 &   6.27    &   0.61    &   \cite{38}. \cite{37}   \\
NGC2310 &    S0                 &   24  &   0.41    &   66.9    &   1.32    &    0.7    &   \cite{59}  \\
NGC2787 &   S0-a    &   21.4    &   0.33    &   668 &   7.03    &   0.53    &   \cite{11}   \\
NGC2964 &   Sbc &   18  &   0.48    &   428 &   4.86    &   0.74    &   \cite{24}. \cite{25}   \\
NGC3054 &   Sb  &   27.9    &   0.39    &   180 &   4.57    &   0.63    &   \cite{21}   \\
NGC3115 &    E-S0               &   25.5    &   0.5 &   2370    &   45.8    &    0.81   &   \cite{67}  \\
NGC3203 &    S0-a               &   24.3    &   0.33    &   115 &   7.59    &    0.53   &   \cite{59}  \\
NGC3245 &   S0  &   15.87   &   0.48    &   881 &   9.54    &   0.78    &   \cite{11}   \\
NGC3390 &    Sb                 &   35.8    &   0.51    &   167 &   7.37    &    0.76   &   \cite{59}  \\
NGC3516 &   S0  &   14.21   &   0.19    &   97.5    &   5.8 &   0.31    &   \cite{11}   \\
NGC3941 &   S0  &   20  &   0.39    &   422 &   4.15    &   0.64    &   \cite{49}  \\
NGC3957 &    S0-a               &   21.6    &   0.37    &   204 &   10.2    &    0.59   &   \cite{59}  \\
NGC3998 &   S0  &   16.3    &   0.39    &   1240    &   13.4    &   0.64    &   \cite{49}   \\
NGC4036 &   S0    &   20  &   0.35    &   498 &   6.43    &   0.57    &   \cite{49}   \\
NGC4102 &   SABb    &   19  &   0.57    &   505 &   5.74    &   0.93    &   \cite{24}, \cite{26}   \\
NGC4111 &   S0-a    &   25.4    &   0.57    &   689 &   17  &   0.92    &   \cite{50}   \\
NGC4138 &   S0-a    &   17.4    &   0.31    &   380 &   7.77    &   0.5 &   \cite{51}   \\
NGC4150 &   S0  &   12.4    &   0.57    &   400 &   4.94    &   0.93    &   \cite{38}   \\
NGC4270 &   S0  &   12.1    &   0.67    &   362 &   9.58    &   1.09    &   \cite{52}   \\
NGC4350 &   S0  &   15  &   0.48    &   742 &   7.52    &   0.77    &   \cite{50}  \\
NGC4352 &    S0                 &   17.3    &   0.43    &   46.6    &   2.83    &    0.7    &   \cite{7}    \\
NGC4449 &   IB  &   44.5    &   0.43    &   57.2    &   0.31    &   0.7 &   \cite{56}. \cite{57}   \\
NGC4469 &    S0-a               &   32  &   0.4 &   526 &   31.85   &   0.65    &    \cite{59} \\
NGC4474 &   S0  &   13.9    &   0.49    &   250 &   3.49    &   0.83    &   \cite{7}    \\
NGC4594 &    Sa                 &   62  &   0.32    &   382 &   26.9    &   0.52     &  \cite{65}  \\
NGC4703 &    Sb                 &   12.7    &   0.64    &   424 &   10.7    &    1.03   &   \cite{59}  \\
NGC4710 &    S0-a               &   17.6    &   0.72    &   222 &   2.91    &    1.17   &   \cite{59}  \\
NGC474  &   S0  &   24.1    &   0.69    &   108 &   5.21    &   1.11    &   \cite{6}    \\
NGC4934 &   S0-a    &   6.44    &   0.34    &   122 &   3.89    &   0.55    &   \cite{53}   \\
NGC5529 &    Sc                 &   29.96   &   0.35    &   190 &   3.61    &    0.6    &   \cite{62}  \\
NGC5574 &   E-S0    &   7.9 &   0.75    &   565 &   7.84    &   1.21    &   \cite{7}    \\
NGC5678 &   SABb    &   21  &   0.5 &   226 &   3.04    &   0.8 &   \cite{24}, \cite{27}  \\
NGC5746 &    SABb               &   16.4    &   0.67    &   1140    &   2.38    &    1.3    &   \cite{59}   \\
NGC5869 &    S0                 &   16.4    &   0.74    &   741 &   24.8    &    1.2    &   \cite{7}   \\
NGC615  &   Sb  &   15.1    &   0.55    &   380 &   2.83    &   0.9 &   \cite{21}  \\
NGC6722 &    Sb                 &   22  &   0.39    &   161 &   4.58    &   0.66    &    \cite{59}  \\
NGC6771 &    S0-a               &   18  &   0.54    &   328 &   18.4    &   0.8 &    \cite{59} \\
NGC6925 &   Sbc &   28.2    &   0.42    &   153 &   1.82    &   0.68    &   \cite{21}   \\
NGC7079 &   S0  &   17.1    &   0.42    &   351 &   6.21    &   0.67    &   \cite{54}   \\
NGC7123 &    Sa                 &   17  &   0.53    &   461 &   17.3    &   0.96     &  \cite{59}   \\
NGC7332 &   S0  &   18.7    &   0.59    &   351 &   5.61    &   0.96    &   \cite{55}   \\
NGC7457 &   S0    &   12.7    &   0.62    &   449 &   5.86    &   1.06    &   \cite{11}   \\
NGC7531 &   SABb    &   25  &   0.52    &   192 &   2.78    &   0.84    &   \cite{21}   \\
NGC891  &    Sb                 &   107 &   0.33    &   94.3    &   5.76    &    0.57   &   \cite{62}  \\
 PGC44931   &    Sc                 &   22.65   &   0.36    &   84.9    &   3.34     &  0.58    &    \cite{59} \\
UGC03087    &   S0  &   4.84    &   0.32    &   201 &   1.16    &   0.52    &   \cite{11}   \\
UGC08823    &   S0  &   4.45    &   0.44    &   274 &   3.08    &   0.75    &   \cite{11}   \\
M83 & SBc & 105 & 0.5 & 240 & 1.88 & 0.91 & \cite{79} \\
M94 & Sab & 57.2 & 0.56 & 447 & 2.66 & 0.89 & \cite{79} \\
\hline
\end{longtable}


\end{document}